\documentstyle[sprocl,epsf]{article}

\def\Journal#1#2#3#4{{#1} {\bf #2}, #3 (#4)}

\def\NPB{{\em Nucl. Phys.} B}

\def\be{\begin{equation}}
\def\ee{\end{equation}}
\def\bea{\begin{eqnarray}}
\def\eea{\end{eqnarray}}

\begin{document}
\hspace*{10.0cm}OCHA-PP-140

\title{HEAVY HIGGS PRODUCTIONS AT $\gamma \gamma$ COLLIDERS
\footnote{based on the work in collaboration with
J. Kamoshita, A. Sugamoto and I. Watanabe~\cite{as}.}}

\author{ ERI ASAKAWA }

\address{Department of Physics and
Graduate School of Humanities and Sciences  \\
Ochanomizu University, Otsuka 2-1-1, 
Bunkyo-ku, Tokyo 112-8610, Japan}


\maketitle\abstract{Multi-Higgs doublet models include
more than one neutral Higgs bosons, some of which have 
different properties under a CP transformation.
We examine their productions at $\gamma \gamma$ colliders.
It is found that helicity observations of final top-pairs 
can be powerful tools to distinguish their CP properties,
especially when masses of the Higgs bosons are almost degenerate.}
  
\section{Introduction}

Searches for Higgs boson(s) and precise measurements of their properties
are important to understand the mechanism of the electroweak
symmetry breaking.
Although the standard model (SM) predicts one neutral CP-even Higgs boson,
several extended models of the SM,
such as supersymmetric SM (SSM),
require more than one Higgs bosons. In such models, 
neutral CP-odd Higgs boson(s) ($A$)
as well as neutral CP-even Higgs boson(s) ($H$) appear.
Models like the minimal supersymmetric extension of the SM (MSSM)
which is one of the two-Higgs doublet models
are expected to have almost degenerate masses for H and A
if they are heavy in comparison with $m_Z$.
We examine the possibility of observing these Higgs bosons
and distinguishing their CP properties at $\gamma \gamma$ colliders
in such cases where their mass difference is at most the same
order of their total decay widths.

We show that helicity observations of final top-pairs~\cite{ha}
can be powerful
tools for our purposes in spite of dull peak distributions
of luminosity which are peculiar to $\gamma \gamma$ colliders.
This advantage comes from the difference between final $t_L \bar{t}_L$
and $t_R \bar{t}_R$ processes in
effects of interference between amplitudes.

In this report, we discuss the detection possibility of heavy
neutral Higgs bosons at future $\gamma \gamma$ colliders,
especially concentrating on the MSSM case as an example.

\section{Helicity dependence of amplitudes}
We define spin-zero states decaying into $t_L \bar{t}_L$ 
and $t_R \bar{t}_R$ systems as $|LL>$ and $|RR>$.
These states are interchanged under a CP transformation,
\begin{eqnarray}
{\cal CP} |LL> & = & - |RR> \ , \nonumber \\
{\cal CP} |RR> & = & - |LL> \ .
\label{parity}
\end{eqnarray}
Therefore,
the CP eigenstates are 45$^\circ$ mixtures of $|LL>$ and $|RR>$:
\begin{eqnarray}
{\cal CP} (|LL> - |RR>) & = & + (|LL> - |RR>) \ , \nonumber \\
{\cal CP} (|LL> + |RR>) & = & - (|LL> + |RR>) \ .
\label{eigen}
\end{eqnarray}
The former CP-even eigenstate corresponds to $H$ and
the latter CP-odd eigenstate to $A$.
The CP properties play important roles in the  helicity dependence
of amplitudes.
We here consider the process of $\gamma \gamma \rightarrow
t \bar{t}$. $H$ and $A$ are produced in s-channel via loops of
charged particles. Besides these Higgs production diagrams,
tree diagrams also contribute around Higgs mass poles.
Thus, amplitudes we have to consider are                                       
\begin{equation}
{\cal M}_{\gamma\gamma \rightarrow t\bar{t}}%
^{\lambda \bar{\lambda}} \simeq
  {\cal M}_H^{\lambda \bar{\lambda}}
+ {\cal M}_A^{\lambda \bar{\lambda}}
+ {\cal M}_{cont}^{\lambda \bar{\lambda}} \ ,
\label{amp}
\end{equation}
denoting the helicities of $t$ and $\bar{t}$  as $\lambda$ and
$\bar{\lambda}$, respectively, and 
${\cal M}_H^{\lambda \bar{\lambda}}$,
${\cal M}_A^{\lambda \bar{\lambda}}$ represent amplitudes of 
H, A productions,
${\cal M}_{cont}^{\lambda \bar{\lambda}}$ of tree diagrams,
where the subscript `cont' means features
of the tree diagrams, that is, continuum dependence on center-of-mass energy, 
$\sqrt{s}_{\gamma \gamma}$.

In Table 1, we show relative signs of helicity amplitudes,
where ${\cal M}_H$ stands for ${\cal M}_H^{RR}$ and
${\cal M}_A$ for ${\cal M}_A^{RR}$ 
at $\gamma_+ \gamma_+$ collisions.
The amplitudes for the tree diagrams are given as follows;
\begin{eqnarray}
{\cal M}_{cont}^{RR}=-8 \pi \alpha Q_t^2
\frac{m_t (1+\beta_t)}{E_t (1-\beta_t^2 \cos^2 \theta)}
\end{eqnarray}
\begin{eqnarray}
{\cal M}_{cont}^{LL}=-8 \pi \alpha Q_t^2
\frac{m_t (1-\beta_t)}{E_t (1-\beta_t^2 \cos^2 \theta)},
\end{eqnarray}
where $Q_t$, $m_t$, $E_t$ and $\beta_t$ are
charge, mass, energy and beta factor of top quark 
in the center-of-mass frame. The absolute value of the amplitude
for $t_R \bar{t}_R$ is larger than that for $t_L \bar{t}_L$ due to
the difference in factors $(1\pm\beta_t)$.

First, we consider the case of initial $\gamma_+ \gamma_+$.
The minus sign of the $H$ amplitude for $\gamma_+ \gamma_+
\rightarrow t_L \bar{t}_L$ comes from the CP transformation for
the final top-pair. As for the $A$ amplitude, 
the odd contribution is 
cancelled by another CP-odd factor from the coupling 
between $A$ and the top-pair.
We can see from the analysis that $H$-$A$ interference is 
cancelled if the helicities of final top-pairs are not
observed.

On the other hand, the CP transformation for initial photons
do not induce a minus sign. 
In the same way as final top-pairs,
polarized initial photons lead to non-vanishing 
$H$-$A$ interference.

\begin{table}[t]
\caption{Helicity dependence of the amplitudes for
$\gamma \gamma \rightarrow t \bar{t}$\label{tab:exp}}
\vspace{0.4cm}
\begin{center}
\begin{tabular}{|c|c|c|}
\hline
&
${\large t_R \bar{t}_R}$ &
${\large t_L \bar{t}_L}$
\\ 
\hline
{\large $\gamma_+ \gamma_+$}&
\begin{minipage}{1.0in}
\begin{center}
${\cal M}_{cont}^{RR}$\\
${\cal M_H} $\\
${\cal M_A}$
\end{center}
\end{minipage}
&
\begin{minipage}{1.0in}
\begin{center}
${\cal M}_{cont}^{LL}$\\
$-{\cal M_H} $\\
${\cal M_A}$
\end{center}
\end{minipage}
\\ \hline
{\large $\gamma_- \gamma_-$}&
\begin{minipage}{1.0in}
\begin{center}
${\cal -M}_{cont}^{LL}$\\
${\cal M_H} $\\
$-{\cal M_A}$
\end{center}
\end{minipage}
&
\begin{minipage}{1.0in}
\begin{center}
$-{\cal M}_{cont}^{RR}$\\
$-{\cal M_H} $\\
$-{\cal M_A}$ 
\end{center}
\end{minipage}
\\ 
\hline
\end{tabular}
\end{center}
\end{table}

\section{Interference between amplitudes}
We give numerical estimates of the cross sections for the above
processes~\cite{ga}. 
To simplify the Higgs sector, we concentrate on MSSM 
having two Higgs doublets.
Then we are able to perform definite numerical
estimations by fixing a few MSSM parameters, without losing
the essence which are applicable for the more
complicated models. We have three states of 
the neutral Higgs bosons in MSSM, the light
Higgs $h$, the heavy Higgs $H$ and the pseudoscalar Higgs $A$.
The former two are CP-even, and the last CP-odd.
In our analyses, we parametrize the Higgs sectors by two
parameters, the mass of $A$, $m_A$, and the ratio of the vacuum
expectation values of two Higgs doublets, $\tan\beta$.
In MSSM, the value of the mass of $h$, $m_h$, is close to $m_A$,
if $A$ is lighter than $Z$ boson.
On the other hand, as $m_A$ becomes heavier, the mass of $H$, $m_H$
approaches to $m_A$, and $m_h$ saturates the upper bound of roughly
150 GeV.
We now focus on the heavy $A$ case where both $H$ and $A$ can decay into 
$t \bar{t}$.
\begin{figure}[t]
\begin{center}
\epsfxsize=15cm
\epsfysize=11cm
\epsffile{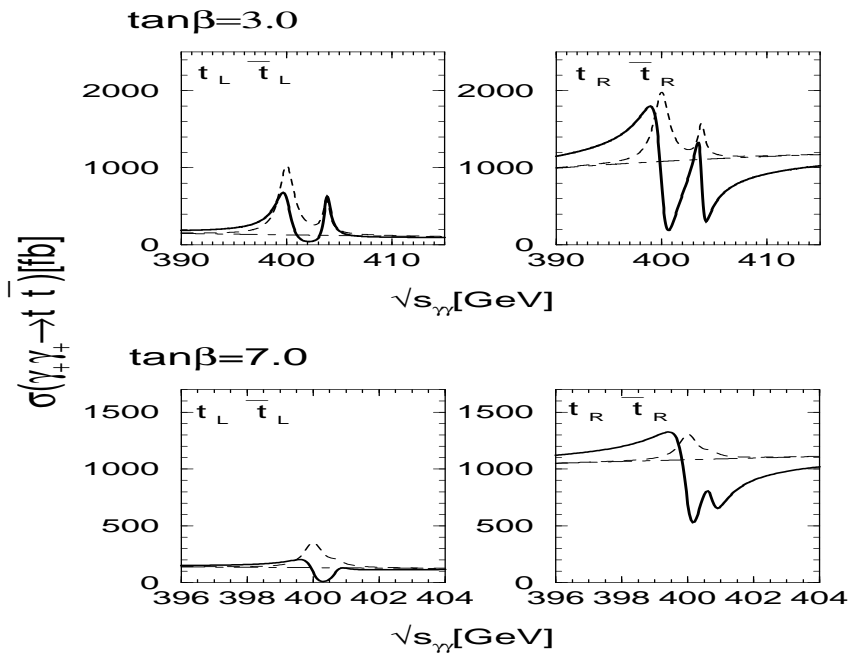}
\vspace*{-13mm}
\caption{$\sqrt{s}_{\gamma \gamma}$ dependence 
of bare cross sections for $\gamma_+ \gamma_+ \rightarrow t \bar{t}$}
\label{diagrams}
\end{center}
\end{figure}

Fig.1 shows $\sqrt{s}_{\gamma \gamma}$ dependence 
of cross sections in the case of $\tan \beta = 3.0$ and 
$7.0$.
We took values of MSSM parameters as
$M_2 = 500$GeV, $\mu = -500$GeV, $M_{\tilde{f}}=1000$GeV
which give masses to charged 
SUSY particles in loops
and evade large contribution of
them owing to their heaviness.
We fixed the mass of $A$ to $400$GeV, and other parameters,
such as the mass of $H$, total decay widths of $H$ and $A$, 
are calculated by HDECAY ~\cite{hdecay}.
The dotted curves represent the cross sections without interference
and the dot-dashed ones that without any Higgs production.
The effects of interference can be clearly observed by
comparing dotted curves with solid ones 
which are complete cross sections
including interference.

Here, we notice on the difference of interference effects 
between two processes, $\gamma_+ \gamma_+ \rightarrow
t_L \bar{t}_L$ and $\gamma_+ \gamma_+ \rightarrow t_R \bar{t}_R$.
The interference terms are proportinal to 
real parts of the product of two in
${\cal M}_{cont}$, ${\cal M}_H$ and ${\cal M}_A$.
Since the absolute values of the continuum amplitudes
are larger than that of $H$, $A$ amplitudes
except for near mass poles of two Higgs bosons,
the effects from the interference terms 
between ${\cal M}_{cont}$ and
${\cal M}_H$, ${\cal M}_A$ 
contribute significantly.
Signs of these terms are changed as
$\sqrt{s}_{\gamma \gamma}$ goes over 
the mass poles of $H$, $A$. 
The estimation is shown in Table 2.
It can be seen in $t_R \bar{t}_R$ case 
that constructive effects are expected below $m_A$, 
destructive above $m_H$.
In $t_L \bar{t}_L$ case, between the masses of two,
the terms contribute destructive. 
These features can be seen in Fig.1 even in larger
$\tan \beta$ where branching ratios of Higgs bosons 
decaying into top-pairs are smaller due to $\tan \beta$ dependence
of their couplings.

\begin{table}[t]
\caption{ $\sqrt{s}_{\gamma \gamma}$ dependence 
of signs of interference terms 
between $M_{cont}$ and $M_{H,A}$\label{tab:exp}}
\vspace{0.4cm}
\begin{center}
\begin{tabular}{|c|c|c|c|c|}
\hline
\multicolumn{2}{|c|}{ }&
$\sqrt{s}_{\gamma \gamma}< m_A$&
$m_A<\sqrt{s}_{\gamma \gamma}< m_H$&
$m_H<\sqrt{s}_{\gamma \gamma}$
\\
\hline
& A--cont & {\bf +} & {\bf --} & {\bf --}
\\
{\large $t_R \bar{t}_R$}
& H--cont & {\bf +} & {\bf +} & {\bf --}
\\
\hline
& A--cont & {\bf +} & {\bf --} & {\bf --}
\\
{\large $t_L \bar{t}_L$}
& H--cont & {\bf --} & {\bf --} & {\bf +}
\\ \hline
\end{tabular}
\end{center}
\end{table} 

\section{Convoluted cross sections}
Next, we must consider $\sqrt{s}_{\gamma \gamma}$ dependence
of luminosity. The convoluted cross sections are obtained 
as follows;
\begin{equation}
\sigma^*(\sqrt{s}_{ee}) = \int d \sqrt{s}_{\gamma \gamma}\frac{1}{L}
\frac{d L(\sqrt{s}_{\gamma \gamma})}{d \sqrt{s}_{\gamma \gamma}}
\sigma (\sqrt{s}_{\gamma \gamma}),
\label{eq:murnf}
\end{equation}
where $\sqrt{s}_{ee}$ is center-of-mass energy of original electrons
and $L$ the $\gamma \gamma$ luminosity.
\begin{figure}[t]
\begin{center}
\epsfxsize=15cm
\epsfysize=6cm
\epsffile{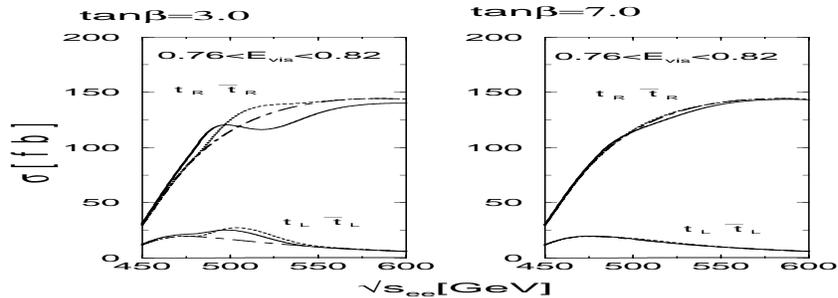}
\vspace*{-13mm}
\caption{$\sqrt{s}_{ee}$ dependence 
of convoluted cross sections for 
$\gamma_+ \gamma_+ \rightarrow t \bar{t}$}
\label{diagrams}
\end{center}
\end{figure}
Fig.2 shows the convoluted cross sections for 
$\tan \beta = 3.0$, $7.0$
after considering the cuts for the visible energies in the detector,
$E_{vis}$.
Though the remarkable effects of interference for bare cross sections
are unfortunately smeared, 
constructive effects below the $H$ and $A$ masses and destructive
effects above the masses remain in $t_R \bar{t}_R$ case.
When $\tan \beta$ becomes larger or $t_L \bar{t}_L$ cases
are considerd, very high luminosity
and careful analyses seem to be required in order to observe
the interference effects.
However, they certainly have information of the existence of 
two Higgs bosons which are scalar and pseudoscalar.
\\
\section{Summary}
We have discussed heavy Higgs productions at $\gamma \gamma$
colliders.
It has been found that the helicity observations of final
top-pairs can be powerful strategy for heavy Higgs detections
at $\gamma \gamma$ colliders, as long as the Higgs bosons have
enough branching ratios of decaying into top-pairs.

A similar method can be performed more clearly for muon
colliders via direct $\mu^+ \mu^-$ annihilation into
$H$ or $A$, with the muon beam polarization.
It will be discussed in a separate paper.
\\
\section*{Acknowledgments}
We would like to thank K. Hagiwara for valuable discussions
and comments.
This work is supported in part by the Grant-in-Aid for
Scientific Research (No. 11640262) and the Grant-in-Aid
for Scientific Research on Priority Areas (No. 11127205)
from the Ministry of Education, Science and Culture, Japan.
\\


\section*{References}
\begin{thebibliography}{99}
\bibitem{as}
E. Asakawa, J. Kamoshita, A. Sugamoto and I. Watanabe,
Ochanomizu University preprint, OCHA-PP-130, in preparation.
\\
\bibitem{ha}K. Hagiwara, H. Murayama and I. Watanabe,
\Journal{\NPB}{367}{257}{1991}.
\\
\bibitem{ga}J.F. Gunion, H.E. Haber, G. Kane and S. Dawson, 
{\em `Higgs Hunter's Guide'} (Addison-Wesley 
Publishing Company, 1990).
\\
\bibitem{hdecay}
A. Djouadi, J. Kalinowski and M. Spira,
\Journal{\em Comput. Phys. Commun.}{108}{56}{1998}.
\end{thebibliography}
\end{document}